\documentclass[a4paper,twocolumn,english]{revtex4-1}
\usepackage[T1]{fontenc}
\usepackage[latin9]{inputenc}
\usepackage{babel}
\usepackage{bm}
\usepackage{amsmath}
\usepackage{amssymb}
\usepackage{graphicx}
\usepackage{wasysym}
\usepackage{esint}
\usepackage[unicode=true,pdfusetitle,
 bookmarks=true,bookmarksnumbered=false,bookmarksopen=false,
 breaklinks=false,pdfborder={0 0 1},backref=false,colorlinks=false]
 {hyperref}
\usepackage{breakurl}



\begin{document}

\title{Pseudo chiral anomaly in zigzag graphene ribbons}

\author{Chang-An Li}

\affiliation{Institute of Natural Sciences, Westlake Institute for Advanced Study, Hangzhou, Zhejiang, China}

\affiliation{School of Science, Westlake University, Hangzhou, Zhejiang, China}

\affiliation{Department of Physics, The University of Hong Kong, Pokfulam Road,
Hong Kong, China}

\date{\today }

\begin{abstract}
As the three-dimensional analogs of graphene, Weyl semimetals display
signatures of chiral anomaly which arises from charge pumping between
the lowest chiral Landau levels of the Weyl nodes in the presence
of parallel electric and magnetic fields. In this work, we study the
pseudo chiral anomaly and its transport signatures in graphene ribbon
with zigzag edges. Here \textquotedblleft pseudo\textquotedblright{}
refers to the case where the inverse of width of zigzag graphene ribbon
plays the same role as magnetic field in three-dimensional Weyl semimetals.
The valley chiral bands in zigzag graphene ribbons can be introduced
by edge potentials, giving rise to the nonconservation of chiral current,
i.e., pseudo chiral anomaly, in the presence of a longitudinal electric
field. Further numerical results reveal that pseudo magnetoconductivity
of zigzag graphene ribbons is positive and has a nearly quadratic
dependence on the pseudofield, which is regarded as the transport
signature of pseudo chiral anomaly.

\end{abstract}
\maketitle

\section{Introduction}

Graphene, a celebrated semimetal with nearly vanishing band gaps,
holds massless Dirac fermions in two dimensions (2D) \cite{Castro09rmp,Novoselov05nature}.
The Dirac fermions in graphene behave in unusual ways leading to remarkable
transport properties such as anomalous integer quantum Hall effect
\cite{Novoselov05nature,ZhangYB05nat,JiangZ07ssc,Novoselov07science},
minimum conductivity \cite{Ando02jpsj,Tworzdlo06prl,Tan07prl}, and
Klein tunneling \cite{Katsnelson,Stander09prl}. As analogs of graphene
in 3D, topological Dirac and Weyl semimetals have attracted intense
research interest over the past years \cite{Hosur13phys,Armitage18rmp}.
The unique topological nature of Dirac and Weyl semimetals indicates
novel properties such as Fermi arc states on their surface \cite{WanX11prb,HuangSM15natcomm,XuS15science}.
Also the chiral anomaly, a well-known phenomenon in quantum field
theory \cite{Peskin}, may show its existence in topological Dirac
and Weyl semimetals through characteristic transport properties such
as chiral magnetic effect and negative magnetoresistance \cite{Kim13prl,HuangXC15prx,LiC15nc,LiH16nc,Liang18prx}.

Graphene shares several characters with Weyl semimetals. First, the
lower-energy physics near the graphene nodes (band touching points
$K$ and $K'$) and Weyl nodes in Weyl semimetals are governed by
Weyl equation. The graphene nodes, which are dubbed Dirac points,
nevertheless, are actually described by Weyl Hamiltonian in 2D \cite{YangSY16spin}.
Second, topological invariants can be defined from the Weyl nodes
as well as graphene nodes. For Weyl semimetals, the integration of
Berry curvature on a simple closed surface enclosing a Weyl node gives
the topological charge (magnetic monopole charge) that works as a
topological invariant. Similarly, the graphene nodes give rise to
a topological charge $N_{1}=\frac{1}{4\pi i}\mathrm{Tr}[\hat{S}\varoint d\ell\mathcal{H}^{-1}\partial_{\ell}\mathcal{H}]$
\cite{Volovik} where $\hat{S}$ is the chiral (sublattice) symmetry
operator and $\mathcal{H}$ is the effective Hamiltonian of graphene
nodes. In addition, the Weyl nodes, as well as graphene nodes, come
in pairs with opposite topological charges in the whole Brillouin
zone according to the no-go theorem \cite{Nielson81npb1,Nielson81npb2}.
Third, there are surface states connecting two nodes with opposite
topological charges both in graphene and Weyl semimetals. The Fermi
arcs surface states in Weyl semimetals are associated with a well-defined
Chern number \cite{Hosur13phys}, and the surface (edge) states in
graphene ribbons with zigzag edges correspond to winding numbers \cite{Ryu02prl}.

The similarities between graphene and Weyl semimetals naturally lead
to a question: Is it possible to reveal some properties of Weyl semimetals,
such as chiral anomaly and its related transport signatures, on graphene
basis? The chiral anomaly, also called Adler--Bell--Jackiw anomaly,
is a quantum anomaly which states the violation of chiral symmetry
of a classical action in the corresponding quantized theory \cite{Peskin,Adler69pr,Bell69nca}.
The chiral anomaly in Weyl semimetals is readily understood in terms
of Landau levels in a crystal \cite{Nielsen83pl}, and it promises
novel transport properties including the negative magnetoresistance
\cite{Nielsen83pl,DTSon13prb,Burkov14prl} and chiral magnetic effect
\cite{Fukushima08prd,LiQ16np}. Generally, chiral anomaly only exists
in $(1+1)$D and $(3+1)$D \cite{Peskin}, and it should be absent
in graphene. Whereas the graphene ribbons, synthesized with atomical
precision \cite{Cai10nature,Ruffieux16nature}, serve as quasi-1D
systems. Besides endowed with many interesting properties and applications
\cite{Han07prl,Tao11np,Wang16NC}, graphene ribbons are even designed to mimic
topological quantum phases in 1D \cite{Groening18nature,Rizzo18nature}.
Indeed, the spectrum of graphene ribbon with zigzag edges closely
resembles the Landau level of Weyl semimetals in the presence of a
magnetic field (consider the dispersive direction) \cite{Hosur13phys,Lu15prb}.
Thereby the answer to the aforementioned question may be affirmative
\cite{LiC172D}.

In the following, we extend the chiral anomaly to graphene ribbon
with zigzag edges in a phenomenological way, and investigate its transport
related signatures numerically. To this end, we first revisit the
topological nature of surface states in zigzag graphene ribbons in
terms of Su-Schriffer-Heeger (SSH) model, then introduce truly chiral
bands by confining zigzag graphene ribbons with edge potentials. Further
application of an external electric field parallel to the ribbon drives
the chiral bands with opposite chirality to give rise to chiral anomaly.
This pseudo chiral anomaly in graphene lead to the finite-size conductivity
bearing a positive dependence on effective magnetic field $\mathcal{B}\equiv1/W$
with $W$ the width of ribbon. With the help of numerical calculations,
we show that the pseudo magnetoconductivity indeed has a nearly quadratic
dependence on effective magnetic field $\mathcal{B}$.

\section{Model and methods}

\subsection{Zigzag graphene ribbon as Su-Schriffer-Heeger model}

The existence of surface states in graphene is determined by the geometry
of its edges \cite{Nakada96prb,Brey06prb}. Typically, the graphene
ribbons with zigzag (beared) edges exhibit localized surface states
while the ribbons with armchair edges have no such surface states.
It is possible to characterize surface states of graphene ribbons
with different edge structures, namely, zigzag, beared, and armchair
edges, even arbitrary edge geometry, topologically \cite{Ryu02prl,Hatsugai11jpcs,Delplace11prb,CaoT17prl,Rhim18prb}.

\begin{figure}[h]
\includegraphics[width=9cm]{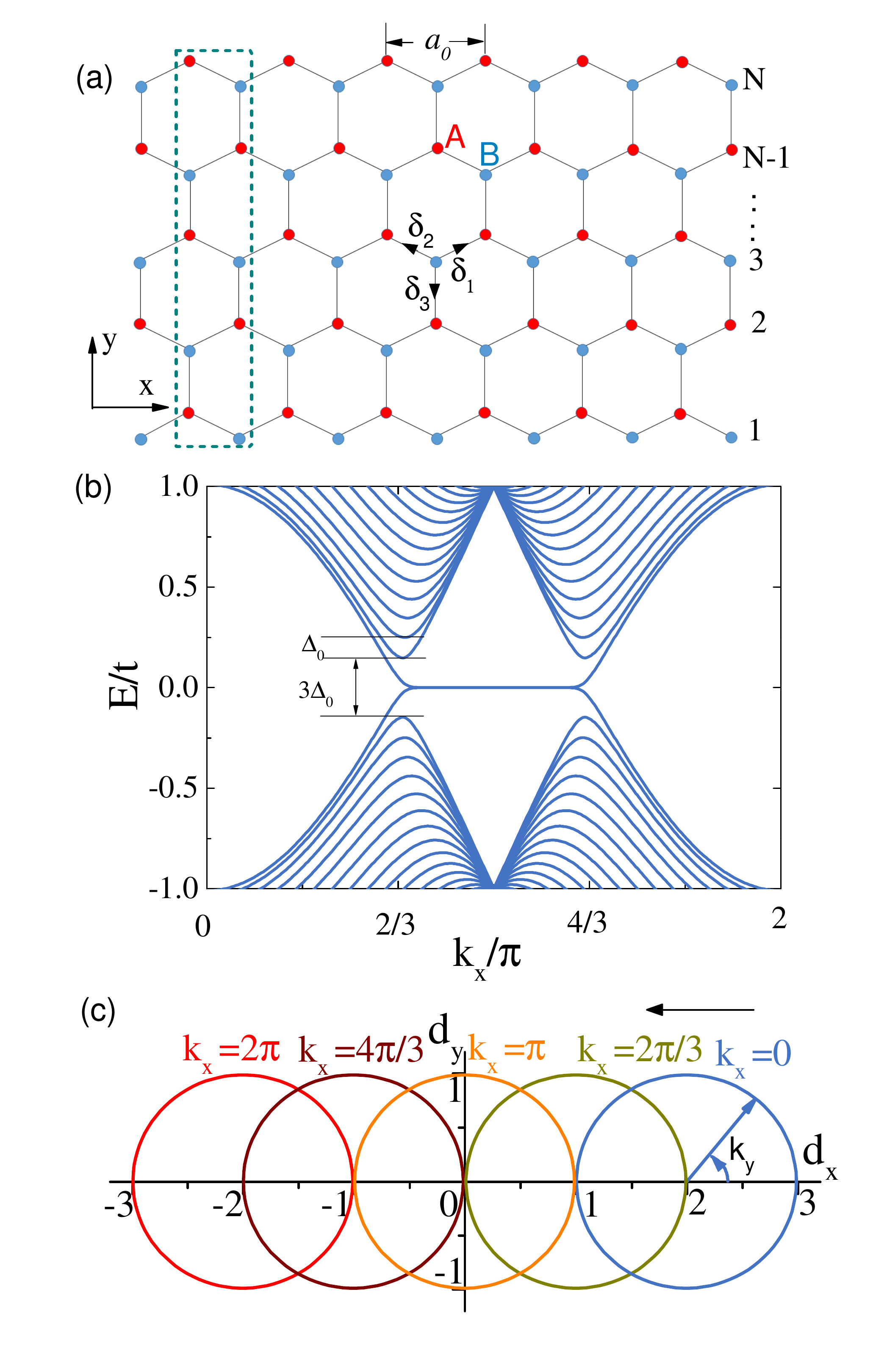}

\caption{The surface states in graphene ribbon with zigzag edges. (a) Scheme
of graphene ribbon with zigzag edges. $N$ is the number of zigzag
lines. Here $a_{0}=\sqrt{3}a$. (b) The spectrum of graphene ribbon
with zigzag edges. (c) The end point of vector $\hat{{\bf d}}({\bf k})$
forms unit circles as changing $k_{y}$. Here $k_{y}$ works as a
polar angle.}
\label{fig:SSH}
\end{figure}

We find it is more concise and straightforward to understand the surface
states by identifying the graphene ribbon as an SSH model in the reduced
1D parameter space. SSH model describes spinless electrons sitting
in a 1D dimer chain, and it has been extensively studied as a prototype
for associated topological properties \cite{SSH79prl,Shen2017,Asboth2015}.
In the following we make a simple connection between the SSH model
and graphene ribbons. Graphene has a honeycomb lattice structure as
shown in Fig. \ref{fig:SSH}(a). The three nearest-neighbor vectors
in real space are denoted as $\bm{\delta}_{1}=\frac{a}{2}(\sqrt{3},1),\ \ \bm{\delta}_{2}=\frac{a}{2}(-\sqrt{3},1),\ \ \bm{\delta}_{3}=a(0,-1)$
with $a$ the carbon-carbon distance. Considering only nearest-neighbor
hopping, the tight-binding Hamiltonian for electrons in graphene reads
\begin{equation}
H=t\sum_{\bm{R}_{B}}\sum_{\ell=1}^{3}C_{\bm{R}_{B}+\bm{\delta}_{\ell}}^{\dagger}C_{\bm{R}_{B}}+h.c.,
\end{equation}
where $C_{\bm{R}}^{\dagger}$($C_{\bm{R}}$) creates (annihilates)
an electron at site $\bm{R}$. Here the spin degree of freedom is
suppressed and $t$ is the nearest-neighbor hopping energy. In the
following, we first focus our attention on graphene ribbon with zigzag
edges and then analyze ribbons with beared and armchair edges. After
performing partial Fourier transformation in $x$ direction, the zigzag
ribbon is reduced to 1D fictitious chain along $y$ direction, in
which each unit cell contains two sites. The hopping energy in and
between unit cells of the chain is different, i.e., $2t\cos(\sqrt{3}ak_{x}/2)$
for intracell hopping and $t$ for intercell hopping. Therefore, the
reduced zigzag ribbon is identical to SSH model of dimerized 1D lattice.
The spectrum of graphene ribbons with zigzag edges obtained in this
way is shown in Fig. \ref{fig:SSH}(b), in which the flat band surface
states connect $K$ and $K'$ valleys. Imposing periodic boundary
condition along $y$ direction further, the system is described by
the Hamiltonian
\begin{equation}
h_{k_{x}}^{Z}(k_{y})=[2\cos\frac{\sqrt{3}k_{x}a}{2}+\cos\frac{3k_{y}a}{2}]\sigma_{x}+\sin\frac{3k_{y}a}{2}\sigma_{y}\equiv\bm{d}(\bm{k})\cdot\mathbf{\bm{\sigma}},\label{eq:ZigzagH}
\end{equation}
where $\mathbf{\bm{\sigma}\equiv}(\sigma_{x},\sigma_{y})$ is the
Pauli matrix vector, and $\bm{d}(\bm{k})=(2\cos(\sqrt{3}k_{x}a/2)+\cos(3k_{y}a/2),\sin(3k_{y}a/2))$.
The wave numbers are normalized with corresponding primitive translation
vectors, and the energy is scaled by hopping energy $t$. Note that
the primitive lattice spacings in $x$ and $y$ directions are $\sqrt{3}a$
and $3a/2$, respectively. Diagonalizing the Hamiltonian of Eq. \ref{eq:ZigzagH},
we obtain the spectrum that is consistent with the very one of zigzag
graphene \cite{Rhim18prb,Waka99prb}. It should be noted that Eq.
\ref{eq:ZigzagH} is exactly the SSH Hamiltonian if we redefine $d_{x}(k_{y})=v+w\cos k_{y}$
and $d_{y}(k_{y})=w\sin k_{y}$ with $v\equiv2\cos(k_{x}/2)$ and
$w\equiv1$ \cite{Shen2017,Asboth2015}. The topology of SSH model
is characterized by the winding number defined as $\bm{\nu}=\frac{1}{2\pi i}\oint q^{-1}dq$
with $q=v+we^{-ik_{y}}$. Since $\bm{\nu}=1$ for $|v|<|w|$ in SSH
model, the zigzag graphene is topologically nontrivial with wave vector
$k_{x}$ locating in the region $(\frac{2\pi}{3},\frac{4\pi}{3})$.
From the bulk-boundary correspondence, there are surface states connecting
$K$ and $K'$ valleys, as shown in Fig. \ref{fig:SSH}(b). For the
region $k_{x}\in(0,\frac{2\pi}{3})\cup(\frac{4\pi}{3},2\pi)$, the
winding number is zero and no surface states are expected. We can
also interpret the topology of graphene ribbons in terms of the Zak
phase \cite{Zak88prl}, which gives consistent results as winding
numbers.

The vector $\bm{d}(\bm{k})$ provides a more vivid way to explore
the topology of Eq. \ref{eq:ZigzagH}. The path of the end point of
vector $\bm{d}(\bm{k})$, as the wave number $k_{y}$ goes through
the 1D Brillouin zone $[0,2\pi]$, is a unit circle centered at $\left(2\cos(k_{x}/2),0\right)$
on the $d_{x}$-$d_{y}$ plane, as shown in Fig. \ref{fig:SSH}(c).
Increasing $k_{x}$ moves the unit circle left. Once the circle cuts
the origin, it gives Dirac cone. Varying the wave vector $k_{x}$
from $0$ to $2\pi$, the gap of the reduced 1D system will close
at $k_{x}=2\pi/3$ then reopen, and similar process happens again
around $k_{x}=4\pi/3$. This gap close-reopen process may indicate
topological phase transition. Whether or not the system is topologically
nontrivial is determined by the origin is enclosed by the unit circle
or not. Similar analysis can also be applied to graphene ribbon with
beared and armchair edges. For example, the reduced Hamiltonian for
the beared edge case is $h_{k_{x}}^{B}(k_{y})=[1+2\cos(k_{x}/2)\cos k_{y}]\sigma_{x}+2\cos(k_{x}/2)\sin k_{y}\sigma_{y}.$
Following the same argument, there are flat band surface states in
the region $k_{x}\in(0,\frac{2\pi}{3})\cup(\frac{4\pi}{3},2\pi)$.
For the ribbons with armchair edges, there are no such surface states.

\subsection{Constructing valley chiral bands by edge potentials}

We demonstrated in the previous section that the surface states in
the zigzag graphene ribbons can be interpreted by SSH model.In this
section we show how to construct genuinely chiral bands using these
surface states in order to realize pseudo chiral anomaly in graphene
ribbon with zigzag edges.

\begin{figure}[h]
\includegraphics[width=1\linewidth]{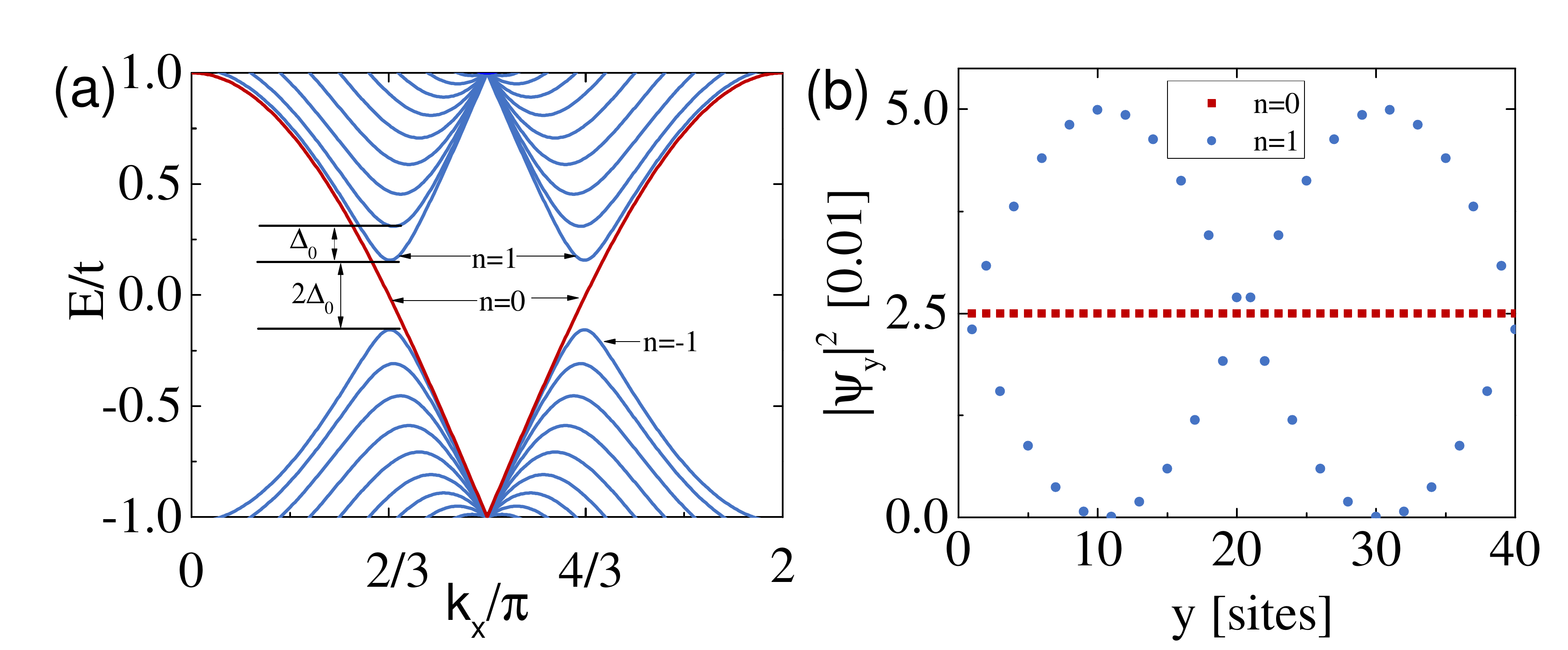}

\caption{Constructing the gapless chiral bands by introducing edge potentials.
(a) The spectrum of graphene ribbon with zigzag edges confined by
edge potentials. The chiral bands are emphasized by red lines. (b)
The wave function for $k_{x}=2\pi/3$ corresponding to panel (a).}
\label{fig:chiralmode}
\end{figure}

The wave functions of surface states at $k_{x}=\pi$ with energy $E=0$
are entirely localized at the edges of the ribbon. Applying edge potentials
$U_{\mathrm{edge}}$ at the outmost sites of two edges, intuitively,
drags the surface states to the states with energy $E=U_{\mathrm{edge}}$
\cite{Yao12prl}. Considering the band structures as shown in Fig.
\ref{fig:SSH}(b), in which the energies of all other bands coincide
at $E=\pm1$ for $k_{x}=\pi$, we choose the confining edge potentials
to be $U_{\mathrm{edge}}=\pm1$. For the case of $U_{\mathrm{edge}}=-1$,
gapless chiral bands, labeled as $n=0$, cross the valleys, as seen
in Fig. \ref{fig:chiralmode}(a). Focusing on the left (right) valley,
modes $n=1,2,\cdots,N$ lie above the $n=0$ chiral band, whereas
modes $n=-1,-2,\cdots,-(N-1)$ lie below the $n=0$ chiral band, and
the gapless chiral band cross the valley with group velocity $v_{F}=-\frac{3}{2}\frac{ta}{\hbar}$
($v_{F}=+\frac{3}{2}\frac{ta}{\hbar}$). The two $n=0$ chiral bands
should be stable, as will be shown, since they originate from the
topologically protected surface states.

The wave numbers $k_{x}=2\pi/3$ and $k_{x}=4\pi/3$ are unique points
at which the system has special energy and wave functions. For $k_{x}=2\pi/3$
and $k_{x}=4\pi/3$, the Hamiltonian for the zigzag graphene ribbons
confined by the edge potentials turns out to be
\begin{equation}
h_{\eta}^{Z}=\left(\begin{array}{ccccc}
U_{\mathrm{edge}} & -\eta & \cdots & 0 & 0\\
-\eta & 0 & \ddots & 0 & 0\\
\vdots & \ddots & \ddots & \ddots & \vdots\\
\vdots & \vdots & -\eta & 0 & -\eta\\
0 & \cdots & 0 & -\eta & U_{\mathrm{edge}}
\end{array}\right)\label{eq:H_2pi/3}
\end{equation}
under the basis $(C_{\eta A1},C_{\eta B1},C_{\eta A2},C_{\eta B2},\cdots,C_{\eta AN},C_{\eta BN})^{T}$.
Here $\eta=-1$ for $k_{x}=2\pi/3$ and $\eta=+1$ for $k_{x}=4\pi/3$,
respectively. For $U_{\mathrm{edge}}=-1,$ the eigen values of Eq.
\ref{eq:H_2pi/3} read

\begin{equation}
E_{\eta,j}=2\cos[j\pi/(2N)],\ \ \ j=1,2,...,2N.
\end{equation}
The zero-energy modes, corresponding to the chiral bands $n=0$, always
exist when $j=N$. By Taylor expansion and elimination of $N$ from
the width $W=(3N/2-1)a$, the mode spacing under the limit $W\gg1$
is

\begin{equation}
\Delta_{0}=3ta\pi/(2W),
\end{equation}
which is proportional to $1/W$. The energy separation of the conduction
and valence bands, as shown in Fig. \ref{fig:chiralmode}(a), is $2\Delta_{0}$
now and different from $3\Delta_{0}$ as in Fig. \ref{fig:SSH}(b)
\cite{Rycerz07np}. The corresponding eigenvectors for $n=0$ chiral
bands are

\begin{equation}
\psi_{\eta}(y)=\frac{1}{\sqrt{2N}}(1,-\eta,-1,\eta,1,-\eta,\cdots)^{T}.
\end{equation}
It is interesting to note from Fig. \ref{fig:chiralmode}(b) that
the wave functions of the $n=0$ chiral bands are not localized at
edges but distribute equally on each site along $y$ direction (it
is even true at $k_{x}\neq\eta$), which is different from the case
where the bulk gap is opened \cite{Yao12prl}.

\section{Results}

\subsection{Pseudo chiral anomaly in zigzag graphene ribbons}

The spectrum in Fig. \ref{fig:chiralmode}(a) is very similar to that
of Weyl fermions in the presence of a strong magnetic field \cite{Hosur13phys},
and it provides the possibility to realize chiral anomaly in graphene
ribbons. Crucially, the chiral bands have definite chiralities just
as the chiral Landau levels of Weyl fermions. Suppose the temperature
is at zero and only the chiral bands are relevant, thus the system
is in the so called `quantum limit' regime. Applying an external electric
field with strength $\bm{E}$ along the ribbon will drive electrons
flowing from one valley to the other. The change rate of charge carrier
density at one valley $\eta$($=\pm$) in the quasi-1D system is $\frac{d\rho_{\eta}}{dt}=\frac{\eta}{W}\frac{1}{2\pi}\frac{dk_{x}}{dt},$
and the momentum vector will obey the equation $\frac{dk_{x}}{dt}=\frac{e\bm{E}}{\hbar}$.
Therefore, we have $\frac{d\rho_{\eta}}{dt}=\eta\frac{e}{h}\frac{\bm{E}}{W}$.
Similar to chiral anomaly in Weyl semimetals, here we can also define
an equation for divergence free current $j_{5}^{\mu}$ as
\begin{equation}
\partial_{\mu}j_{5}^{\mu}=\frac{\partial\rho_{+}}{\partial t}-\frac{\partial\rho_{-}}{\partial t}=\frac{2e}{h}\bm{E}\mathcal{B},\label{eq:chiral anomaly graphene}
\end{equation}
where $\mu=0,1$, $\partial_{0}\equiv\partial_{t}$ and $\partial_{1}\equiv\partial_{x}$,
and $\mathcal{B}\equiv1/W$. Physically, Eq. \ref{eq:chiral anomaly graphene}
states nonconservation of the chiral charge in the presence of electric
field and effective magnetic field $\mathcal{B}$. Thus, by the same
mechanism that is operative in 3D Weyl semimetals \cite{Nielsen83pl,Pikulin16prx},
the chiral anomaly can be generalized to quasi-1D graphene ribbons.
Comparing with the current conservation equation of Weyl semimetals,
$\mathcal{B}$ here takes the role of a magnetic field. This analogy
also makes sense in terms of physical picture of Landau levels: In
graphene ribbons, the energy band spacing $\Delta_{0}$ decreases
with decreasing the effective magnetic field $\mathcal{B}$, i.e.,
$\Delta_{0}\propto\mathcal{B}$; and likewise the Landau level separation
also decreases with decreasing the real magnetic field strength. It
should be noted here we mainly focus on the orbital effect of a magnetic
field imposed on electrons. The difference between the effective magnetic
field $\mathcal{B}$ and a real magnetic field, however, lies in several
aspects: First, the effective magnetic field $\mathcal{B}$ preserves
time reversal symmetry; second, the unit of effective magnetic field
$\mathcal{B}$ is not as well-defined as the real magnetic field;
third, the effective magnetic field $\mathcal{B}$ has no Zeeman effect.
For these reasons above, the chiral anomaly presented here in graphene
ribbons is therefore referred to as \textit{pseudo chiral anomaly}.

\subsection{Transport signatures of the pseudo chiral anomaly}

\begin{figure}[h]
\includegraphics[width=1\linewidth]{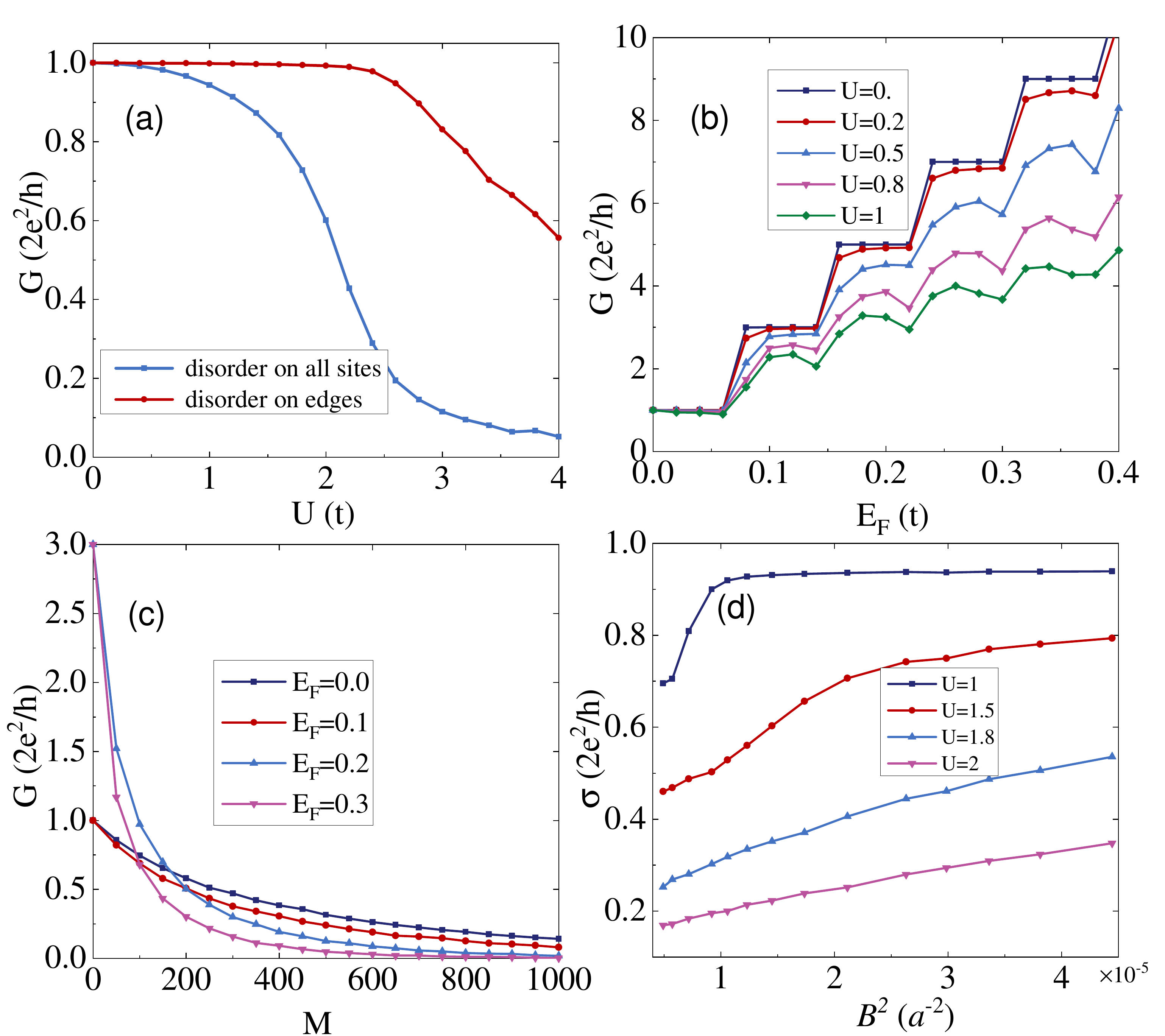}

\caption{Transport of chiral bands under the effect of onsite disorders. (a)
Conductance as functions of disorder strength $U$ with $E_{F}=0.01$,
$M=34$ and $N=40$. (b) Conductance as functions of $E_{F}$ with
$M=34$ and $N=40$. (c) Conductance as functions of length $M$ with
$U=1$ and $N=20$. (d) The conductivity as function of effective
magnetic field square $\mathcal{B}^{2}$ with $E_{F}=0.01$, and in
this plot we take $N$ from 100 to 301. Here 5000 disorder configurations
are taken for all.}
\label{fig:chiralanomaly_onsite}
\end{figure}

Due to the interplay of chiral magnetic effect and intervalley scattering,
the pseudo magnetoconductivity $\sigma(\mathcal{B})$ should be positive
and have quadratic dependence on $\mathcal{B}$ in the diffusive transport
regime \cite{LiC172D,LiQ16npA}, which is regarded as the signature
of pseudo chiral anomaly in graphene. In this section, we try to reveal
this anomalous scaling behavior of finite-size conductivity numerically.
From Fig. \ref{fig:chiralmode}(a), the left valley locates at $k_{-}=2\pi/3$
and the right valley locates at $k_{+}=4\pi/3$. The rather large
momentum difference $\Delta k=2\pi/3$ between the two valleys indicates
that the transport properties depend on impurity ranges heavily. As
mentioned above, the signature of chiral anomaly will only be revealed
with intervalley scattering. Thereby two typical kinds of disorders,
namely, onsite disorders and Gaussian type disorders, are considered.
The onsite disorders are considered by independently adding every
lattice site a potential drawn from uniform distribution $[-U/2,U/2]$
with $U$ the disorder strength. For Gaussian type disorders uniformly
distributed in the 2D real space, the potential on a lattice site
with position $\bm{R}$ has the form \cite{Wakabayashi07prl}

\begin{equation}
V(\bm{R})=\sum_{\bm{r}_{i}}u\exp(-|\bm{R}-\bm{r}_{i}|^{2}/d^{2}),
\end{equation}
where $d$ is the disorder range and $\bm{r}_{i}$ is the disorder
position. Here $u$ is the disorder strength uniformly distributed
within $[-u_{m},u_{m}]$ constrained by the normalization condition
\begin{equation}
\sum_{\bm{R}\in\mathrm{full\ space}}u_{m}\exp(-|\bm{R}|^{2}/d^{2})/(\sqrt{3}/2)=u_{0}.
\end{equation}

The Landau-B$\ddot{\mathrm{u}}$ttiker formalism provides an efficient
way to calculate the transport properties of electrons. Here we consider
a two-terminal setup with disordered central region connected to two
clean leads. With the help of recursive Green's function techniques,
the conductance of two-terminal device can be evaluated as

\begin{equation}
G=\frac{2e^{2}}{h}Tr\left[\Gamma_{L}G^{r}\Gamma_{R}G^{a}\right],
\end{equation}
where $\Gamma_{L,R}$ are the line-width functions coupling to left
lead and right lead, respectively, and $G^{r}(G^{a})$ is the retarded
(advanced) Green's function of the disordered region \cite{Dattabook}.
Note that the conductivity meets $\sigma=G$ for a square sheet \cite{Tworzdlo06prl,Lewenkopf08prb}.
Here the width $W=(3N/2-1)a$ and length $L=\sqrt{3}Ma$ with $M$
the number of defined unit cells in Fig. \ref{fig:SSH}(a).

\begin{figure}[h]
\includegraphics[width=1\linewidth]{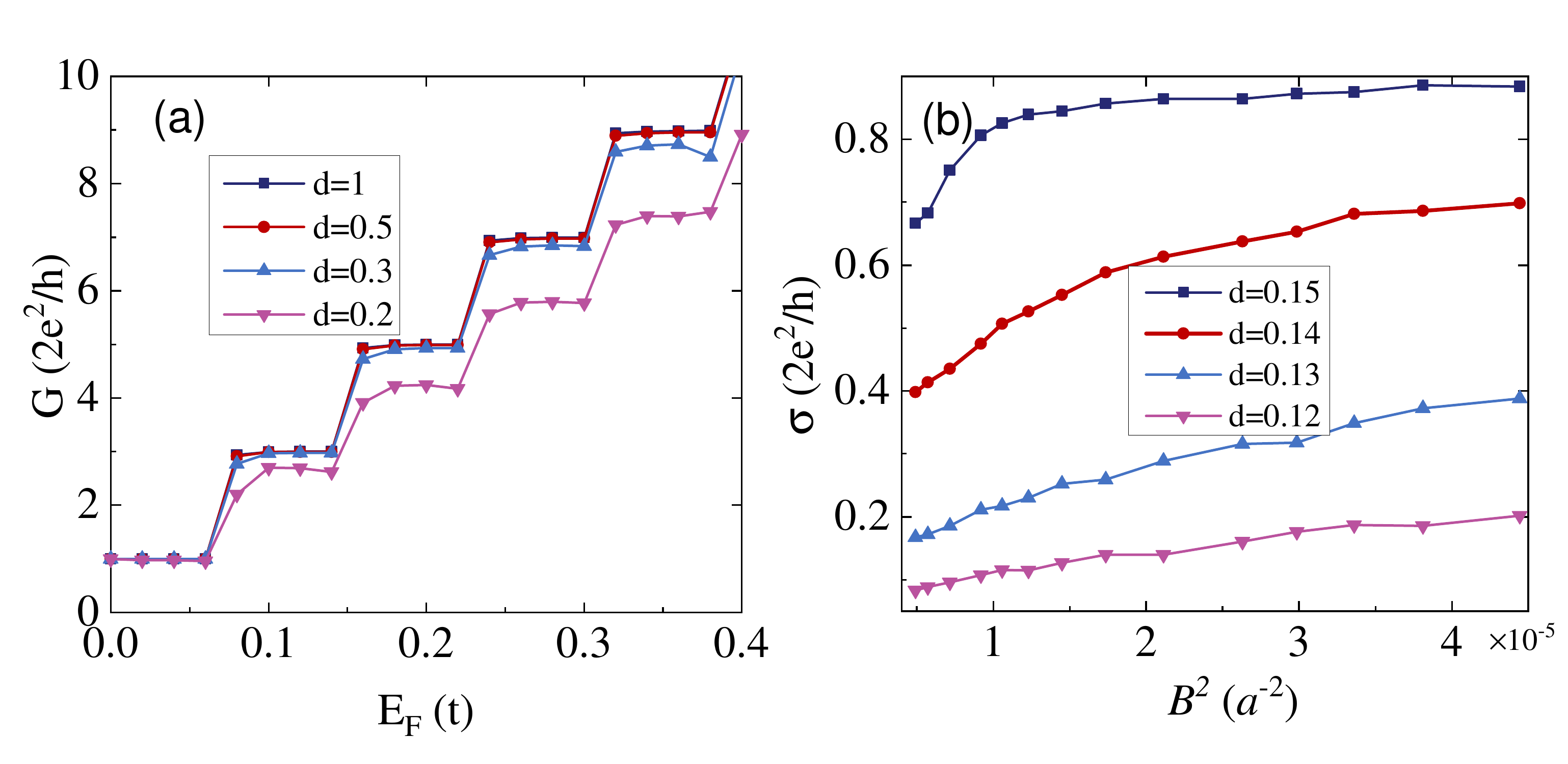}

\caption{Transport of chiral bands under the effect of Gaussian type disorder.
(a) Conductance as functions of $E_{F}$ with $M=34$ and $N=40$.
(b) The conductivity as function of effective magnetic field square
$\mathcal{B}^{2}$ with $E_{F}=0.01$, $u_{0}=1$. Here 5000 disorder
configurations are taken for all.}
\label{fig:chiralanomaly_Gaussian}
\end{figure}

Let us now consider the transport properties of the $n=0$ chiral
bands. For a clean sample, the relevant chiral bands should underlie
a quantized conductance $2e^{2}/h$. To verify the validity of chiral
bands induced by edge potentials, the onsite disorders are added at
the outmost sites of the two edges first. As seen in Fig. \ref{fig:chiralanomaly_onsite}(a),
the quantized conductance from chiral bands persists until edge disorder
strength reaches critical strength $U_{c}\simeq2.5$. Therefore, the
mechanism of confining edge potentials still works as long as the
disorder strength does not exceed the critical value $U_{c}$. While
for the case where disorders are on all sites, the conductance decreases
quickly as shown by blue line in Fig. \ref{fig:chiralanomaly_onsite}(a).
Seen from Fig. \ref{fig:chiralanomaly_onsite}(b) (also in Fig. \ref{fig:chiralanomaly_Gaussian}(a)),
the $n=0$ chiral bands are more stable than other bands: The conductance
plateau from $n=0$ chiral bands is nearly unchanged as increasing
the disorder strength. The scaling behavior of conductance with respect
to sample length also shows the robustness of the $n=0$ chiral bands
(see Fig. \ref{fig:chiralanomaly_onsite}(c)).

For onsite disorders, the pseudo magnetoconductivity as function of
$\mathcal{B}^{2}$ with different disorder strengths is presented
in Fig. \ref{fig:chiralanomaly_onsite}(d): Focusing on the curve
with disorder strength $U=1$, the conductivity drops from a constant
value linearly with respect to $\mathcal{B}^{2}$ as soon as the system
is in laterally diffusive regime where the width exceeds mean free
path; with appropriate disorder strength, the pseudo magnetoconductivity
has a nearly quadratic dependence on effective magnetic field $\mathcal{B}$
and it is treated as the transport signature of pseudo chiral anomaly.
Here the Fermi energy shifts away from zero slightly to avoid strong
fluctuations \cite{Adam07pnas}, and we keep the assumption that the
Fermi energy only crosses the two $n=0$ chiral bands. For Gaussian
disorders, we fix the disorder strength at $u_{0}=1$ and vary the
interacting range $d$. For a shorter interacting range $d$, the
intervalley scattering is relatively stronger. It is shown from Fig.
\ref{fig:chiralanomaly_Gaussian}(b) that the pseudo magnetoconductivity
shares similar behavior as in the onsite disorder case, i.e., it shows
nearly quadratic scaling behavior on effective magnetic field $\mathcal{B}$.
The mechanism for this \textquoteleft anomalous\textquoteright{} scaling
feature is the joint effort of finite size confinement and chiral
magnetic effect. The electron change rate is proportional to $1/W$
in the quasi-1D system, then small chemical potential difference induced
from the balance between intervalley scattering and chiral magnetic
effect also contains a factor $1/W$ \cite{LiC172D}. Thus the current
density is proportional to $1/W^{2}$. While note that the previous
results show that the conductivity of graphene near the charge neutral
point has weak dependence on size \cite{Lewenkopf08prb,Cresti07prb},
and it reaches at a minimal value at order of $e^{2}/h$ \cite{Tan07prl,Terres16nc}.
Comparison of Fig. \ref{fig:chiralanomaly_onsite}(d) and \ref{fig:chiralanomaly_Gaussian}(b)
indicates that transport properties of chiral bands exhibit little
dependence on the disorder types but much on disorder strengths.

\section{Discussion }

Varying the termination direction away from zigzag case will gradually
weak the pseudo chiral anomaly and finally it disappears at armchair
case. Thus we mainly focus on zigzag type graphene ribbons, and it
is mostly relevant for experiments. One possible concern about the
analogy between graphene and Weyl semimetals is that the stability
of graphene nodes, unlike Weyl nodes, needs to be protected by inversion
symmetry \cite{BernevigBook}. Consider staggered potentials $U_{A}=\Delta$
for $A$ sites and $U_{B}=-\Delta$ for $B$ sites universally in
the graphene lattice. At this moment, the bulk bands are gaped with
amount $2\Delta$. The winding number is not well-defined now due
to the existence of mass term in the bulk Dirac Hamiltonian. However,
there are still flat bands with energy $E=\pm\Delta$ \cite{Delplace11prb,Yao12prl}.
After applying proper edge potentials, the chiral bands crossing two
valleys persist as in Fig. \ref{fig:chiralmode}(a), thereby the physics
of pseudo chiral anomaly remains. Besides, the magnetic order at zigzag
edges of graphene ribbons may have influence on the bulk states \cite{Fujita96jpsj}.
While the band gap opened by the magnetic order is negligible as long
as the zigzag ribbons wider than 8 nanometers \cite{Magda14nature},
which is satisfied in our considerations, i.e., $N=40\sim300$ and
$W=8.5\sim64$ nm. Note the strong intervalley scattering can lead
to Anderson localization. While generally the typical size scales
regarding localization properties in 2D systems are very large, thus
the 2D electron localization mechanism may have little influence on
our quasi-1D system \cite{Fan14prb,Bardarson07prl}.

\section{Conclusions}

In conclusion, we show that pseudo chiral anomaly may exist in graphene
ribbon with zigzag edges. We first show that the zigzag graphene ribbon
can be mapped to SSH model, and then we turn the flat band surface
states of zigzag graphene ribbons to chiral bands, based on which
the pseudo chiral anomaly can be realized. This pseudo chiral anomaly
underlies anomalous scaling behavior of pseudo magnetoconductivity
on the effective magnetic field $\mathcal{B}$. Further numerical
calculations on the transport properties indicates the pseudo magnetoconductivity
has a positive and nearly quadratic dependence on $\mathcal{B}$.
This finding provides graphene as an alternative platform to explore
the purely quantum chiral anomaly in solids.
\begin{acknowledgments}
I would like to thank Alex Weststr$\ddot{\mathrm{o}}$m for helpful
discussions.
\end{acknowledgments}


\begin{thebibliography}{0}%
\makeatletter
\providecommand \@ifxundefined [1]{%
 \@ifx{#1\undefined}
}%
\providecommand \@ifnum [1]{%
 \ifnum #1\expandafter \@firstoftwo
 \else \expandafter \@secondoftwo
 \fi
}%
\providecommand \@ifx [1]{%
 \ifx #1\expandafter \@firstoftwo
 \else \expandafter \@secondoftwo
 \fi
}%
\providecommand \natexlab [1]{#1}%
\providecommand \enquote  [1]{``#1''}%
\providecommand \bibnamefont  [1]{#1}%
\providecommand \bibfnamefont [1]{#1}%
\providecommand \citenamefont [1]{#1}%
\providecommand \href@noop [0]{\@secondoftwo}%
\providecommand \href [0]{\begingroup \@sanitize@url \@href}%
\providecommand \@href[1]{\@@startlink{#1}\@@href}%
\providecommand \@@href[1]{\endgroup#1\@@endlink}%
\providecommand \@sanitize@url [0]{\catcode `\\12\catcode `\$12\catcode
  `\&12\catcode `\#12\catcode `\^12\catcode `\_12\catcode `\%12\relax}%
\providecommand \@@startlink[1]{}%
\providecommand \@@endlink[0]{}%
\providecommand \url  [0]{\begingroup\@sanitize@url \@url }%
\providecommand \@url [1]{\endgroup\@href {#1}{\urlprefix }}%
\providecommand \urlprefix  [0]{URL }%
\providecommand \Eprint [0]{\href }%
\providecommand \doibase [0]{http://dx.doi.org/}%
\providecommand \selectlanguage [0]{\@gobble}%
\providecommand \bibinfo  [0]{\@secondoftwo}%
\providecommand \bibfield  [0]{\@secondoftwo}%
\providecommand \translation [1]{[#1]}%
\providecommand \BibitemOpen [0]{}%
\providecommand \bibitemStop [0]{}%
\providecommand \bibitemNoStop [0]{.\EOS\space}%
\providecommand \EOS [0]{\spacefactor3000\relax}%
\providecommand \BibitemShut  [1]{\csname bibitem#1\endcsname}%
\let\auto@bib@innerbib\@empty
\end{thebibliography}%


\begin{thebibliography}{10}


\bibitem{Castro09rmp}A. H. Castro Neto, F. Guinea, N. M. R. Peres,
K. S. Novoselov, and A. K. Geim, \textit{The electronic properties
of graphene}, Rev. Mod. Phys. \textbf{81}, 109 (2009).

\bibitem{Novoselov05nature}K. S. Novoselov, A. K. Geim, S. V. Morozov,
D. Jiang, M. I. Katsnelson, I. V. Grigorieva, S. V. Dubonos, and A.
A. Firsov, \textit{Two-dimensional gas of massless Dirac fermions
in graphene}, Nature \textbf{438}, 197 (2005).

\bibitem{ZhangYB05nat}Y. Zhang, Y.-W. Tan, H. L. Stormer, and P.
Kim, \textit{Experimental observation of the quantum Hall effect and
Berry's phase in graphene}, Nature \textbf{438}, 201 (2005).

\bibitem{JiangZ07ssc}Z. Jiang and Y. Zhang and Y.-W. Tan and H.L.
Stormer, and P. Kim, \textit{Quantum Hall effect in graphene}, Solid
State Commun. \textbf{143}, 14 (2007).

\bibitem{Novoselov07science}K. S. Novoselov, Z. Jiang, Y. Zhang,
S. V. Morozov, H. L. Stormer, U. Zeitler, J. C. Maan, G. S. Boebinger,
P. Kim, and A. K. Geim, \textit{Room-temperature quantum Hall effect
in graphene,} Science \textbf{315}, 1379 (2007).

\bibitem{Ando02jpsj}T. Ando, Y. Zheng, and H. Suzuura, \textit{Dynamical
conductivity and zero-mode anomaly in honeycomb lattices}, J. Phys.
Soc. Jpn. \textbf{71}, 1318 (2002).

\bibitem{Tworzdlo06prl}J. Tworzydlo, B. Trauzettel, M. Titov, A.
Rycerz, and C. W. J. Beenakker, \textit{Sub-Poissonian shot noise
in graphene}, Phys. Rev. Lett. \textbf{96}, 246802 (2006).

\bibitem{Tan07prl}Y.-W. Tan, Y. Zhang, K. Bolotin, Y. Zhao, S. Adam,
E. H. Hwang, S. Das Sarma, H. L. Stormer, and P. Kim, \textit{Measurement
of scattering rate and minimum conductivity in graphene}, Phys. Rev.
Lett. \textbf{99}, 246803 (2007).

\bibitem{Katsnelson}M. I. Katsnelson, K. S. Novoselov and A. K. Geim,
\textit{Chiral tunneling and the Klein paradox in graphene}, Nat.
Phys. \textbf{2}, 620 (2006).

\bibitem{Stander09prl} N. Stander, B. Huard, and D. Goldhaber-Gordon,
\textit{Evidence for Klein tunneling in graphene p-n junctions,
}Phys. Rev. Lett. \textbf{102}, 026807 (2009).

\bibitem{Hosur13phys}P. Hosur and X. L. Qi, \textit{Recent developments
in transport phenomena in Weyl semimetals}, C. R. Phys. \textbf{14},
857 (2013).

\bibitem{Armitage18rmp}N. P. Armitage, E. J. Mele, and A. Vishwanath,
\textit{Weyl and Dirac semimetals in three-dimensional solids}, Rev.
Mod. Phys. \textbf{90}, 015001 (2018).

\bibitem{WanX11prb}X. Wan, A. M. Turner, A. Vishwanath, and S. Y.
Savrasov, \textit{Topological semimetal and Fermi-arc surface states
in the electronic structure of pyrochlore iridates}, Phys. Rev. B
\textbf{83}, 205101 (2011).

\bibitem{HuangSM15natcomm}S.-M. Huang, S.-Y. Xu, I. Belopolski, C.-C.
Lee, G. Chang, B. Wang, N. Alidoust, G. Bian, M. Neupane, C. Zhang,
S. Jia, A. Bansil, H. Lin, and M. Z. Hasan, \textit{A Weyl Fermion
semimetal with surface Fermi arcs in the transition metal monopnictide
TaAs class}, Nat. Commun. \textbf{6}, 7373 (2015).

\bibitem{XuS15science}S.-Y. Xu, I. Belopolski, N. Alidoust, M. Neupane,
G. Bian, C. Zhang, R. Sankar, G. Chang, Z. Yuan, C.- C. Lee, S.-M.
Huang, H. Zheng, J. Ma, D. S. Sanchez, B. Wang, A. Bansil, F. Chou,
P. P. Shibayev, H. Lin, S. Jia, and M. Z. Hasan, \textit{Discovery
of a Weyl fermion semimetal and topological Fermi arcs}, Science \textbf{349},
613 (2015).

\bibitem{Peskin}M. E. Peskin and D. V. Schroeder, \textit{An introduction
to quantum field theory } (Westview Press Inc., 1995).

\bibitem{Kim13prl}H.-J. Kim, K.-S. Kim, J.-F. Wang, M. Sasaki, N.
Satoh, A. Ohnishi, M. Kitaura, M. Yang, and L. Li, \textit{Dirac versus
Weyl fermions in topological insulators: Adler-Bell-Jackiw anomaly
in transport phenomena}, Phys. Rev. Lett. \textbf{111}, 246603 (2013).

\bibitem{HuangXC15prx}X. Huang, L. Zhao, Y. Long, P. Wang, D. Chen,
Z. Yang, H. Liang, M. Xue, H. Weng, Z. Fang, X. Dai, and G. Chen,
\textit{Observation of the chiral-anomaly-induced negative magnetoresistance
in 3D Weyl semimetal TaAs}, Phys. Rev. X \textbf{5}, 031023 (2015).

\bibitem{LiC15nc}C.-Z. Li, L.-X. Wang, H. Liu, J. Wang, Z.-M. Liao,
and D.-P. Yu, \textit{Giant negative magnetoresistance induced by
the chiral anomaly in individual $Cd_{3}As_{2}$ nanowires}, Nat.
Commun. \textbf{6}, 10137 (2015).

\bibitem{LiH16nc}H. Li, H. He, H.-Z. Lu, H. Zhang, H. Liu, R. Ma,
Z. Fan, S.-Q. Shen, and J. Wang, \textit{Negative magnetoresistance
in Dirac semimetal $Cd_{3}As_{2}$}, Nat. Commun. \textbf{7}, 10301
(2016).

\bibitem{Liang18prx}S. Liang, J. Lin, S. Kushwaha, J. Xing, N. Ni,
R. J. Cava, and N. P. Ong, \textit{Experimental tests of the chiral
anomaly magnetoresistance in the Dirac-Weyl semimetals $Na_{3}Bi$
and GdPtBi}, Phys. Rev. X \textbf{8}, 031002 (2018).

\bibitem{YangSY16spin}S. A. Yang, \textit{Dirac and Weyl materials:
fundamental aspects and some spintronics applications}, SPIN \textbf{6},
1640003 (2016).

\bibitem{Volovik}G. E. Volovik, \textit{The Universe in a Helium
Droplet } (Clarendon Press, 2003).

\bibitem{Nielson81npb1} H. Nielsen and M. Ninomiya, \textit{Absence
of neutrinos on a lattice: (I). Proof by homotopy theory}, Nucl. Phys.
B \textbf{185}, 20 (1981).

\bibitem{Nielson81npb2} H. Nielsen and M. Ninomiya, \textit{Absence
of neutrinos on a lattice: (II). Intuitive topological proof}, Nucl.
Phys. B \textbf{193}, 173 (1981).

\bibitem{Ryu02prl}S. Ryu and Y. Hatsugai, \textit{Topological origin
of zero-energy edge states in particle-hole symmetric systems}, Phys.
Rev. Lett. \textbf{89}, 077002 (2002).

\bibitem{Adler69pr} S. L. Adler, \textit{Axial-vector vertex in spinor
electrodynamics}, Phys. Rev. \textbf{177}, 2426 (1969).

\bibitem{Bell69nca}J. S. Bell and R. Jackiw, \textit{A PCAC puzzle:
$\pi^{0}\rightarrow\gamma\gamma$ in the $\sigma$-model}, Nuovo Cimento
A \textbf{60}, 47 (1969).

\bibitem{Nielsen83pl} H. Nielsen and M. Ninomiya, \textit{The Adler-Bell-Jackiw
anomaly and Weyl fermions in a crystal}, Phys. Lett. B \textbf{130},
389 (1983).

\bibitem{DTSon13prb} D. T. Son and B. Z. Spivak, \textit{Chiral anomaly
and classical negative magnetoresistance of Weyl metals}, Phys. Rev.
B \textbf{88}, 104412 (2013).

\bibitem{Burkov14prl} A. A. Burkov, \textit{Chiral anomaly and diffusive
magnetotransport in Weyl metals}, Phys. Rev. Lett. \textbf{113}, 247203
(2014).

\bibitem{Fukushima08prd} K. Fukushima, D. E. Kharzeev, and H. J.
Warringa, \textit{Chiral magnetic effect}, Phys. Rev. D \textbf{78},
074033 (2008).

\bibitem{LiQ16np} Q. Li, D. E. Kharzeev, C. Zhang, Y. Huang, I. Pletikosic,
A. V. Fedorov, R. D. Zhong, J. A. Schneeloch, G. D. Gu, and T. Valla,\textit{Chiral
magnetic effect in $ZrTe_{5}$}, Nat. Phys. \textbf{12}, 550 (2016).

\bibitem{Cai10nature} J. Cai, P. Ruffieux, R. Jaafar, M. Bieri, T.
Braun, S. Blankenburg, M. Muoth, A. P. Seitsonen, M. Saleh, X. Feng,
K. Mullen, and R. Fasel, \textit{Atomically precise bottom-up fabrication
of graphene nanoribbons}, Nature \textbf{466}, 470 (2010).

\bibitem{Ruffieux16nature} P. Ruffieux, S. Wang, B. Yang, C. S$\acute{a}$nchez-S$\acute{a}$nchez,J.
Liu, T. Dienel, L. Talirz, P. Shinde, C. A. Pignedoli, D. Passerone,
T. Dumslaff, X. Feng, K. M$\ddot{u}$llen, and R. Fasel, \textit{On-surface
synthesis of graphene nanoribbons with zigzag edge topology}, Nature
\textbf{531}, 489 (2016).

\bibitem{Han07prl} M. Y. Han, B. $\ddot{O}$yilmaz, Y. Zhang, and P. Kim,
\textit{Energy band-gap engineering of graphene nanoribbons}, Phys.
Rev. Lett. \textbf{98}, 206805 (2007).

\bibitem{Tao11np}C. Tao, L. Jiao, O. V. Yazyev, Y.-C. Chen, J. Feng,
X. Zhang, R. B. Capaz, J. M. Tour, A. Zettl, S. G. Louie, H. Dai,
and M. F. Crommie, \textit{Spatially resolving edge states of chiral
graphene nanoribbons}, Nat. Phys. \textbf{7}, 616 (2011).

\bibitem{Wang16NC}S. Wang, L.Talirz, C. A. Pignedoli, X. Feng, K.
M$\ddot{u}$llen, R. Fasel, and P. Ruffieux, \textit{Giant edge state
splitting at atomically precise graphene zigzag edges}, Nat. Commun.
\textbf{7}, 11507 (2016).

\bibitem{Groening18nature} O. Gr$\ddot{o}$ning, S. Wang, X. Yao,
C. A. Pignedoli, G. Borin Barin, C. Daniels, A. Cupo, V. Meunier,
X. Feng, A. Narita, K. M$\ddot{u}$llen, P. Ruffieux, and R. Fasel,
\textit{Engineering of robust topological quantum phases in graphene
nanoribbons}, Nature \textbf{560}, 209 (2018).

\bibitem{Rizzo18nature} D. J. Rizzo, G. Veber, T. Cao, C. Bronner,
T. Chen, F. Zhao, H. Rodriguez, S. G. Louie, M. F. Crommie, and F.
R. Fischer, \textit{Topological band engineering of graphene nanoribbons},
Nature \textbf{560}, 204 (2018).

\bibitem{Lu15prb} H.-Z. Lu, S.-B. Zhang, and S.-Q. Shen, \textit{High-field
magnetoconductivity of topological semimetals with short-range potential},
Phys. Rev. B \textbf{92}, 045203 (2015).

\bibitem{LiC172D} S.-Q. Shen, C.-A. Li, and Q. Niu, \textit{Chiral
anomaly and anomalous finite-size conductivity in graphene}, 2D Mater.
\textbf{4}, 035014 (2017).

\bibitem{Nakada96prb} K. Nakada, M. Fujita, G. Dresselhaus, and M.
S. Dresselhaus, \textit{Edge state in graphene ribbons: Nanometer
size effect and edge shape dependence}, Phys. Rev. B \textbf{54},
17954 (1996).

\bibitem{Brey06prb} L. Brey and H. A. Fertig, \textit{Electronic
states of graphene nanoribbons studied with the Dirac equation}, Phys.
Rev. B \textbf{73}, 235411 (2006).

\bibitem{Hatsugai11jpcs} Y. Hatsugai, \textit{Topological aspect
of graphene physics}, J. Phys. Conf. Ser. \textbf{334}, 012004 (2011).

\bibitem{Delplace11prb} P. Delplace, D. Ullmo, and G. Montambaux,
\textit{Zak phase and the existence of edge states in graphene}, Phys.
Rev. B \textbf{84}, 195452 (2011).

\bibitem{CaoT17prl} T. Cao, F. Zhao, and S. G. Louie, \textit{Topological
phases in graphene nanoribbons: junction states, spin centers, and
quantum spin chains}, Phys. Rev. Lett. \textbf{119}, 076401 (2017).

\bibitem{Rhim18prb} J.-W. Rhim, J. H. Bardarson, and R.-J. Slager,
\textit{Unified bulk-boundary correspondence for band insulators},
Phys. Rev. B \textbf{97}, 115143 (2018).

\bibitem{SSH79prl} W. P. Su, J. R. Schrieffer, and A. J. Heeger,
\textit{Solitons in polyacetylene}, Phys. Rev. Lett. \textbf{42},
1698 (1979).

\bibitem{Shen2017} S.-Q. Shen, \textit{Topological Insultaors: Dirac
equation in condensed matter, 2nd ed.} (Springer, Singapore, 2017).

\bibitem{Asboth2015} J. K. Asb$\acute{o}$th, L. Oroszl$\acute{a}$ny,
and A. P$\acute{a}$lyi, \textit{A short course on topological insulators:
band structure topology and edge states in one and two dimensions}
(Springer-Verlag, Berlin, 2015).

\bibitem{Waka99prb} K. Wakabayashi, M. Fujita, H. Ajiki, and M. Sigrist,
\textit{Electronic and magnetic properties of nanographite ribbons},
Phys. Rev. B \textbf{59}, 8271 (1999).

\bibitem{Zak88prl} J. Zak, \textit{Berry's phase for energy bands
in solids}, Phys. Rev. Lett. \textbf{62}, 2747 (1989).

\bibitem{Yao12prl} W. Yao, S. A. Yang, and Q. Niu, \textit{Edge states
in graphene: From gapped flat-band to gapless chiral modes}, Phys.
Rev. Lett. \textbf{102}, 096801 (2009).

\bibitem{Rycerz07np} A. Rycerz, J. Tworzydlo, and C. W. J. Beenakker,
\textit{Valley filter and valley valve in graphene}, Nat. Phys. \textbf{3},
172 (2007).

\bibitem{Pikulin16prx} D. I. Pikulin, A. Chen, and M. Franz, \textit{Chiral
anomaly from strain-induced gauge fields in Dirac and Weyl semimetals},
Phys. Rev. X \textbf{6}, 041021 (2016).

\bibitem{LiQ16npA} Q. Li and D. E. Kharzeev, \textit{Chiral magnetic
effect in condensed matter systems}, Nucl. Phys. A \textbf{956}, 107
(2016).

\bibitem{Wakabayashi07prl} K. Wakabayashi, Y. Takane, and M. Sigrist,
\textit{Perfectly conducting channel and universality crossover in
disordered graphene nanoribbons}, Phys. Rev. Lett. \textbf{99}, 036601
(2007).

\bibitem{Dattabook} S. Datta, \textit{Electronic Transport in Mesoscopic
Systems} (Cambridge University Press, 1995).

\bibitem{Lewenkopf08prb} C. H. Lewenkopf, E. R. Mucciolo, and A.
H. Castro Neto, \textit{Numerical studies of conductivity and Fano
factor in disordered graphene}, Phys. Rev. B \textbf{77}, 081410 (2008).

\bibitem{Adam07pnas} S. Adam, E. H. Hwang, V. M. Galitski, and S.
Das Sarma, \textit{A self-consistent theory for graphene transport},
Proc. Natl. Acad. Sci. USA \textbf{104}, 18392 (2007).

\bibitem{Cresti07prb}A. Cresti, G. Grosso, and G. P. Parravicini,
\textit{Numerical study of electronic transport in gated graphene
ribbons}, Phys. Rev. B \textbf{76}, 205433 (2007).

\bibitem{Terres16nc}B. Terr$\acute{\mathrm{e}}$s, L. A. Chizhova, F. Libisch, J. Peiro,
D. Joger, S. Engels, A. Girschik, K. Watanabe, T. Taniguchi, S. V.
Rotkin, J. Burgd$\ddot{o}$rfer, and C. Stampfer, \textit{Size quantization
of Dirac fermions in graphene constrictions}, Nat. Commun. \textbf{7},
151528 (2016).

\bibitem{BernevigBook} B. A. Bernevig and T. L. Hughes, \textit{Topological
insulators and topological superconductors} (Princeton University
Press, NJ,  2013).

\bibitem{Fujita96jpsj} M. Fujita. K. Wakabayashi, K. Nakada and K.
Kusakabe, \textit{Perculiar localized states at zigzag graphite edge},
J. Phys. Soc. Jpn. \textbf{65}, 1920 (1996).

\bibitem{Magda14nature} G. Z. Magda, X. Jin, I. Hagym$\acute{a}$si,
P. Vancs$\acute{\mathrm{o}}$, Z. Osv$\acute{a}$th, P. Nemes-Incze,
C. Hwang, L. P. Bir$\acute{\mathrm{o}}$, and L. Tapaszt$\acute{\mathrm{o}}$,
\textit{Room-temperature magnetic order on zigzag edges of narrow
graphene nanoribbons}, Nature \textbf{514}, 608 (2014).

\bibitem{Fan14prb} Z. Fan, A. Uppstu, and A. Harju, \textit{Anderson
localization in two-dimensional graphene with short-range disorder:
One-parameter scaling and finite-size effects}, Phys. Rev. B \textbf{89},
245422 (2014).

\bibitem{Bardarson07prl} J. H. Bardarson, J. Tworzyd\l o, P. W. Brouwer,
and C. W. J. Beenakker, \textit{One-Parameter Scaling at the Dirac
Point in Graphene}, Phys. Rev. Lett. \textbf{99}, 106801 (2007).
\end{thebibliography}
\end{document}